\documentclass[prl,aps,10pt,twocolumn,superscriptaddress,notitlepage,nofootinbib,nobibnotes,amssymb,amsmath]{revtex4-1}
\usepackage{natbib}
\usepackage{slashed}
\usepackage{mathtools}
\usepackage{bm}
\usepackage{graphicx}

\newcommand{\beqn}{\begin{eqnarray}}
\newcommand{\eeqn}{\end{eqnarray}}
\newcommand{\beqs}{\begin{subequations}}
\newcommand{\eeqs}{\end{subequations}\\[-2mm]\noindent}
\newcommand{\eq}[1]{(\ref{#1})}

\newcommand{\cL}{{\cal L}}

\newcommand{\bs}{\boldsymbol}

\begin{document}

\title{Direct measurement of a beta function and an indirect check of the Schwinger effect near the boundary in Dirac semimetals}

\author{M. N. Chernodub}
\affiliation{Institut Denis Poisson UMR 7013, Universit\'e de Tours, 37200 France}
\affiliation{Laboratory of Physics of Living Matter, Far Eastern Federal University, Sukhanova 8, Vladivostok, 690950, Russia}

\author{Mar\'ia A. H. Vozmediano}
\affiliation{Materials Science Factory,  Instituto de Ciencia de Materiales de Madrid, CSIC, Cantoblanco; 28049 Madrid, Spain.}

\begin{abstract}
The electric field inside typical conductors drops down exponentially with the screening length determined by an intrinsic length scale of the system such as the density of mobile carriers. We show that in a classically conformal system with boundaries, where the intrinsic length scale is absent, the screening of an external electric field is governed by the quantum conformal anomaly associated with the renormalization of the electric charge. The electric field  decays algebraically  with a  fractional power determined by the beta function of the system. We argue that this ``anomalous conformal screening effect'' is an indirect manifestation of the Schwinger pair production in relativistic field theory. We discuss the experimental feasibility of the proposed  phenomenon in Dirac semimetals what would allow a direct experimental access to the beta function.
\end{abstract}

\date{October 7, 2019}

\maketitle

\paragraph{Introduction.-} 
There is no electrostatic field in the bulk of an ideal conductor. The screening of the electric field occurs due to the presence of mobile charge carriers which, under  the external electric field, redistribute themselves inside the conductor and generate an excess of the electric charge density at its boundary. The redistributed charges create its own electric field which compensates the external field inside the conductor~\cite{Ashcroft}.

The static electric field falls down with an exponential law, $E(x) \sim E(0) e^{- x/\lambda}$, as one moves from the conductor boundary at $x=0$ towards its bulk, $x>0$. The screening length $\lambda$ determines the width of a  layer of the redistributed mobile charges near the boundary. 

For example, at high enough temperature $T$ the charge carriers form a classical thermal plasma characterized by the Debye screening length $\lambda_{\mathrm{D}}$. At low temperature, the system enters a quantum regime of a nonrelativistic Fermi gas characterized by the Fermi-Thomas screening length $\lambda_{\mathrm{FT}}$, which is produced by density (rather than thermal) effects. Both length scales are fixed by a dimensionful quantity, the density of the charge carriers $n$ in a solid: 
\beqn
\lambda_{\mathrm{D}} = \sqrt{\frac{\varepsilon_0 k_B T}{n e^2}},
\qquad
\lambda_{\mathrm{FT}} = \sqrt{\frac{\varepsilon_0 \pi^2 \hbar^3}{m e^2  p_F}},
\label{eq:lambdas}
\eeqn
where $\varepsilon_0$ is the vacuum permittivity, $k_B$ is the Boltzmann constant, and $p_F = (2 \pi^2 n)^{1/3} \hbar$ is the Fermi momentum of the particles, which carry the electric charge $e$ and mass $m$.

The exponential screening of a static electric field is the associated to a characteristic length scale such as the Thomas-Fermi wavevector~\cite{Ashcroft}. In the examples given above, the width of the surface layer of the displaced charge carriers is given by the screening scales of the system in the bulk~\eq{eq:lambdas}. In the conformal limit of vanishing density and temperature,  the Fermi liquid phenomenology can not be used and these expressions do not apply.

A physical system is conformal (or scale) invariant at the classical level when all its parameters are dimensionless quantities. The question of the electrostatic screening in a classically conformal system is relevant to a wide class of recently discovered Dirac and Weyl semimetals~\cite{Letal14a,Letal14b,NX14,Xu15,Lvetal15,Xuetal15} whose low-energy properties are described by relativistic massless fermions~\cite{NaMa16,AMV17}. High interest to these materials is motivated by the fact that they exhibit a plethora of exotic quantum effects restricted, until very recently, to fundamental high-energy systems such as extremely hot quark--gluon plasma~\cite{KLetal12}. In particular, the Dirac and Weyl semimetals manifest a diversity of quantum anomalies \cite{Shifman:1988zk} which lead to various anomaly--related transport phenomena~\cite{Karl14}. The axial anomaly~\cite{KK13,XKetal15,Lietal15,ZXetal16} generates -- via the chiral magnetic effect~\cite{ref:CME} -- the experimentally accessible electric current parallel to the axis of a background magnetic field~\cite{LKetal16}. The mixed axial-gravitational anomaly~\cite{Landsteiner:2011cp} leads to a positive magneto-thermoelectric conductance for collinear temperature gradients and magnetic fields~\cite{CCetal14,Getal17}, while the conformal anomaly is suggested to generate -- via the scale magnetic effect~\cite{Chernodub:2016lbo} -- an anomalous thermoelectric current perpendicular to a temperature gradient and the direction of a background magnetic field~\cite{CCV18,ACV19}. In these material systems, as well as in massless QED,  the photon polarization function that encodes the  screening properties acquires a logarithmic dependence on the renormalization scale \cite{IN12,YMetal14,JG14,RJH16,PFV18}.

In this Rapid Communication we show that the screening of the electrostatic field in a classically conformal conductor with a boundary, such as Dirac or Weyl semimetal, may occur via the quantum conformal anomaly. The screening may be understood as a boundary effect, with the external electrostatic field screened in the interior of a spatially bounded semimetal. To get rid of potentially non-conformal contributions from thermal and density effects we will consider a system at zero-temperature and zero chemical potential. Notice that a flat boundary does not introduce, by itself, any dimensionful parameter, and therefore, it does not affect the screening length in the bulk.

\paragraph{Conformal electromagnetic edge effects.-} 
About 30 years ago, McAvity and Osborn showed~\cite{McAvity:1990we} that a classical electromagnetic field $F_{\mu\nu} = \partial_\mu A_\nu - \partial_\nu A_\mu$ acting on a bounded quantum system of charged particles, generates a singular electric current near the boundary:
\beqn
J^\mu = -  \frac{2 c \beta_e}{e \hbar} \frac{F^{\mu \nu} n_\nu}{x},
\label{eq:J:edge}
\eeqn
where $n^\mu = (0,{\bs n})$ is the inner normal vector to the edge of the system, and $x >0$ is the spatial distance to the boundary along ${\bs n}$. In Refs.~\cite{Chu:2018ntx,Chu:2018ksb} it was found that the prefactor in Eq.~\eq{eq:J:edge} is given precisely by the beta function:
\beqn
\beta_e \equiv \beta_e(e) = \mu \frac{d e(\mu)}{d \mu},
\label{eq:beta}
\eeqn
which describes the renormalization of the electric charge $e = e(\mu)$ at the energy scale $\mu$. The presence of the beta function shows that the effect~\eq{eq:J:edge} is undoubtedly associated with the conformal anomaly in the interacting system~\cite{ref:Book:renormalization}. Remarkably, the current~\eq{eq:J:edge} does not depend on the particular choice of the reflective boundary conditions~\cite{Chu:2018ksb}.

It is instructive to rewrite Eq.~\eq{eq:J:edge} in the components $J^\mu = (c \rho, {\bs J})$. The background magnetic field ${\bs B}$ induces the electric current
\beqn
{\bs J} = - \frac{2 c \beta_e}{e \hbar} \frac{1}{x} {\bs n} \times {\bs B},
\label{eq:CMEE}
\eeqn
which is normal to the axis of the magnetic field and tangential to the edge of the system. The emergence of the quantum current~\eq{eq:CMEE} may be interpreted as a result of skipping orbits of particles and antiparticles created by the quantum fluctuations near the edge of the system in the background magnetic field~\cite{Chu:2018ntx,Chu:2018ksb}. Althoug the presence of the divergent $1/x$ current might seem unlikely, the existence of the ``conformal magnetic edge effect''~\eq{eq:CMEE} was recently demonstrated in first-principles numerical simulations of scalar quantum electrodynamics~\cite{ref:CMEE}. 

In the presence of a static electric field ${\bs E}$, the conformal anomaly~\eq{eq:J:edge} leads to the accumulation of the electric charge density at the boundary~\cite{Chu:2018ntx}:
\beqn
\rho = - \frac{2 \beta_e}{e \hbar c} \frac{{\bs n} {\bs E}}{x}.
\label{eq:CEEE}
\eeqn
Similarly to the scale electromagnetic effects~\cite{Chernodub:2016lbo}, the conformal magnetic~\eq{eq:CMEE} and electric~\eq{eq:CEEE} edge effects arise at zero temperature and vanishing chemical potentials.

\begin{figure}[!thb]
\begin{center}
\vskip 3mm
\includegraphics[scale=0.8,clip=true]{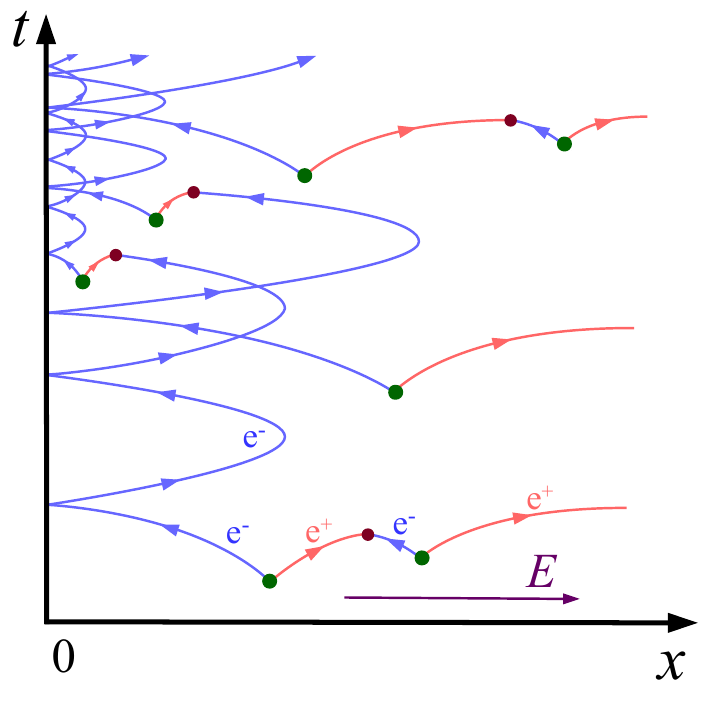}
\end{center}
\vskip -2mm 
\caption{Interpretation of the conformal electric edge effect~\eq{eq:CMEE} as an accumulation of electric charge due to the Schwinger pair production near 
 reflective boundary (placed at $x=0$).}
\label{fig:effect}
\end{figure}

\paragraph{Conformal electric edge effect~\eq{eq:CEEE} and the Schwinger pair production.-} We suggest that the charge accumulation near the boundary~\eq{eq:CEEE} may qualitatively be understood as a consequence of the Schwinger pair production near a reflective boundary, as shown schematically in Fig.~\ref{fig:effect}. In the field theory language, the background electric field leads to the quantum production of particle-antiparticle pairs. The analogous effect in WSMs,  the creation of electron-hole pairs in the presence of a uniform electric field (the Zener effect) has been described in a different context in~\cite{VDM15,Zyuzin18,Abramchuk:2016afc}.
Since the charge carriers are massless, the process of pair production in semimetals proceeds without the prohibitive energy barrier of the usual massive QED where the Schwinger pair production is exponentially suppressed by the electron mass. 

As a pair is created near the boundary, the background electric field accelerates one of the particles towards the boundary and pushes the anti-particle towards the bulk. When the first particle reaches the reflective  boundary, it scatters back into the bulk. Certain part of the reflected particles is subsequently annihilated with particles of an opposite charge that are created in the bulk in later times. Another part of the reflected particles comes back to the reflective boundary which scatters them again into the bulk. As a result of this repetitive processes, the boundary accumulates the electrically charged particles of certain sign which depends on the sign of the product ${\bs n} {\bs E}$ in agreement with Eq.~\eq{eq:CEEE}. An opposite boundary will accumulate the charges of the other sign.  The system reaches an equilibrium when the produced particles form a charged layer which partially screens the external electric field thus stabilizing the vacuum in the bulk. This is the mechanism which unifies the Schwinger pair production and the screening of the electrostatic field by the conformal anomaly.

The particle creation near the boundary lies also in the origin of the conformal magnetic edge effect~\eq{eq:CMEE} which generates the electric current near the boundary in the magnetic field background. This anomalous current originates from the skipping particle--anti-particle orbits in thermal equilibrium~\cite{Chu:2018ksb}. The physical picture of the conformal electric edge effect~\eq{eq:CEEE}, proposed in our article in Fig.~\ref{fig:effect}, is based on the out-of-equilibrium Schwinger process. Thus, although both processes involve similar particle-creation effect near the boundary, the physical pictures behind them are very different.

\paragraph{Anomalous conformal screening effect.-} As an example, consider first the massless QED with $N_f$ species of the Dirac fermions described by the Lagrangian:
\beqn
\cL = - \frac{1}{4} F^{\mu\nu} F_{\mu\nu} + \sum_{a = 1}^{N_f} {\bar \psi}_a i \gamma^\mu D_\mu \psi_a,
\label{eq:L:QED}
\eeqn
where $D_\mu = \partial_\mu - i e A_\mu$ is the covariant derivative expressed via the gauge field $A^\mu = (\phi/c,{\bs A})$. Since in our stationary problem the magnetic field  is absent, ${\bs B} = 0$, we may safely set ${\bs A} = 0$ and use the electrostatic potential~$\phi$ to describe the electric field ${\bs E} = - {\bs\nabla} \phi$.

In order to illustrate the screening effect, let us consider the system~\eq{eq:L:QED} with a single boundary at $x = 0$ in a semi-infinite space $x > 0$. We apply the background electric field normal to the boundary, 
${\bs E} = (E_x,0,0)$ and do not restrict the system in the $yz$ plane. Due to the translational invariance in the $y$ and $z$ directions, the problem becomes one-dimensional with all quantities dependent on the $x$ coordinate only. The relevant Maxwell equation, $\mathrm{div} {\bs E} = \rho$, in one spatial dimension is as follows:
\beqn
\partial_x E_x(x) = \frac{1}{\varepsilon_0}\rho(x)\,,
\label{eq:Maxwell:1d}
\eeqn
where 
\beqn
\varepsilon_0 = \frac{e^2}{4 \pi c \hbar\, \alpha_{\mathrm{QED}}}, 
\label{eq:e0}
\eeqn
is the vacuum permittivity (we use SI units) related to the fine-structure constant $\alpha_{\mathrm{QED}} \simeq 1/137$. 

In a linear-response regime, the electric charge density is determined by the anomalous contribution only~\eq{eq:CEEE}:
\beqn
\rho(x) = - \frac{2 \beta_e}{e c \hbar} \frac{E_x(x)}{x}, \qquad x > 0.
\label{eq:CEEE:1d}
\eeqn
In addition to the anomalous term~\eq{eq:CEEE:1d}, we may also expect the appearance of the thermodynamic contribution to the local charge density
\beqn
\rho_{\mathrm{therm}} = \frac{N_f}{3 \pi^2} \mu^3 + \frac{N_f}{3}  T^2 \mu.
\label{eq:rho:therm}
\eeqn
Here we assumed that the system resides in a thermodynamic equilibrium with the effective local chemical potential $\mu(x) = \phi(x)$ related, via the electrostatic potential $\phi(x)$, to the electric field:
\beqn
E_x(x) = - \partial_x \phi(x).
\label{eq:E:phi}
\eeqn
However, the first term in Eq.~\eq{eq:rho:therm} is vanishing in the linear order while the second term is strictly zero at zero temperature. Therefore we omit the thermodynamic contribution and concentrate on the conformal part~\eq{eq:CEEE:1d} only.

The solutions of Eqs.~\eq{eq:Maxwell:1d}, \eq{eq:CEEE:1d} and \eq{eq:E:phi} for the electrostatic potential $\phi$, the electric field $E$ and the charge density $\rho$ are, respectively, as follows:
\beqs
\beqn
\phi(x) & = & \phi_0 - \frac{C x^{1 - \nu}}{1 - \nu}, 
\label{eq:phi:sol} \\
E_x (x) & = & \frac{C}{x^\nu}, 
\label{eq:E:sol} \\
\rho (x) & = & - \frac{C \varepsilon_0 \nu}{x^{1+\nu}},\qquad
\label{eq:rho:sol} 
\eeqn
\label{eq:sol:minus}
\eeqs
where $C$ and $\phi_0$ are the integration constants and $x>0$. The conformal anomaly screens the electrostatic field in the interior of the semimetal as a polynomial $1/x^\nu$ with the ``conformal screening exponent'',
\beqn
\nu = \frac{2 \beta_e}{e c \hbar \varepsilon_0 },
\label{eq:nu}
\eeqn
determined by the beta function of the electric charge~\eq{eq:beta}. The polynomial screening of the electric field~\eq{eq:E:sol} is natural in the classically conformal regime because the theory has no length parameter to appear in the role of a width of the charge layer at the boundary.

The one-loop beta function of the massless QED~\eq{eq:L:QED},
\beqn
\beta_e = \frac{N_f e^3}{12 \pi^2}.
\label{eq:beta:e}
\eeqn
implies that the conformal screening exponent~\eq{eq:nu} is proportional to the fine structure constant of QED: $\alpha_{\mathrm{QED}}= e^2/(4\pi \epsilon_0 \hbar c)$, a small quantity, $\nu \simeq 1.55 N_f \times 10^{-3}$.

\paragraph{Anomalous conformal screening in semimetals.-} The low-energy physics of chiral relativistic quasiparticle in Dirac and Weyl semimetal is well captured by the Lagrangian (we restore the constants $\hbar$ and $c$):
\beqn
\cL & = & - \frac{1}{4} F^{\mu\nu} F_{\mu\nu} 
\label{eq:L:QED:vF} \\
& & + \sum_{a = 1}^{N_f} {\bar \psi}_a \left[ \gamma^0 \left( i \hbar \frac{\partial}{\partial t}  + e \phi\right) 
+ v_F {\bs \gamma} \left( i \hbar {\bs \nabla} - e {\bs A} \right)\right]\psi_a,
\nonumber 
\eeqn
which is similar to the multi-species massless QED~\eq{eq:L:QED} albeit the appearance of the Fermi velocity $v_F$ in the place of the speed of light in a spatial part of Eq.~\eq{eq:L:QED}.

The derivation of the exponent~\eq{eq:nu} for the semimetal Lagrangian~\eq{eq:L:QED:vF} follows the same steps. Due to the anisotropic dispersion relation originated by the Fermi velocity in \eqref{eq:L:QED:vF}, both the Fermi velocity and the velocity of light (permittivity) are renormalized \cite{KLP02,IN12,RJH16,PFV18}.

The final expression for the conformal screening exponent $\nu$ amounts to replace $\alpha_{\mathrm{QED}}\to \alpha_{\mathrm{WSM}}$:
\beqn
\nu =  \frac{e^2}{6 \pi^2\hbar v_F\varepsilon \varepsilon_0},
\label{eq:nu:vF}
\eeqn
where we also put $N_f =1$. Since the Fermi velocity of typical WSMs is approximately $v_F\sim 10^{-3} c$ and $\varepsilon\sim 10$,  the conformal screening exponent in WSMs is approximately a hundred times bigger than that in the vacuum.

Consider now the semimetal in the form of a slab of a finite length $L$ in the $x$ direction ($0 \leqslant x \leqslant L$). 
We apply the electrostatic potential $\Delta\phi \equiv \phi(x=L) - \phi(x = 0)$ to the opposite boundaries $x=0,L$ of the slab. Without loss of generality we take 
\beqn
\phi(x) = 
\left\{
\begin{array}{ll}
0, \qquad & x = 0 ,\\
\Delta\phi, \qquad & x =L ,
\end{array}
\right.
\label{eq:phi:bc}
\eeqn
The charge density induced by the conformal anomaly~\eq{eq:CEEE} in between the two boundaries is as follows:
\beqn
\rho(x) = - \frac{2 \beta_e}{e c \hbar} \left( \frac{1}{x} - \frac{1}{L - x} \right) E_x(x), \quad 0 < x  < L. \qquad
\label{eq:CEEE:1d:2}
\eeqn

The solutions of Eqs.~\eq{eq:Maxwell:1d}, \eq{eq:E:phi}, \eq{eq:CEEE:1d:2} consistent with the boundary conditions~\eq{eq:phi:bc} on the segment $0< x <L$ are:
\beqs
\beqn
\phi(x) & = & \Delta\phi\, h(\nu) B\left(\frac{x}{L}; 1 - \nu, 1 - \nu\right),  \qquad
\label{eq:phi:sol:2} \\
E_x (x) & = & - \frac{\Delta\phi}{L}  h(\nu) \left[\frac{x}{L} \left(1 - \frac{x}{L} \right)\right]^{-\nu}, 
\label{eq:E:sol:2} \\
\rho (x) & = & \frac{\Delta\phi}{L^2} \varepsilon_0 \nu h(\nu) \left(1-\frac{2x}{L}\right) \left[\frac{x}{L} \left(1 - \frac{x}{L} \right)\right]^{-1-\nu}\!\!,\qquad
\label{eq:rho:sol:2} 
\eeqn
\label{eq:sol:plus}
\eeqs
where 
\beqn
B(z; a, b) = \int_0^z \frac{t^{a-1} dt}{(1-t)^{b-1}}
\eeqn
is the Euler incomplete beta function and
\beqn
h(\nu) = \frac{\Gamma(2- 2\nu)}{\Gamma^2(1-\nu)} \equiv \frac{1}{B\left(1 - \nu, 1 - \nu\right)},
\label{eq:h}
\eeqn
is the normalization coefficient expressed via the gamma function $\Gamma(x)$ and the beta function $B(a, b) \equiv B(1; a, b)$. The coefficient~\eq{eq:h} has an infinite series of poles at $\nu = 1/2 + n$ with $n = 1,2, \dots$ which limits the applicability of the linear-response approximation~\eq{eq:sol:plus} to $|\nu| \ll 3/2$, consistent with phenomenological estimations for~$\nu$.

\begin{figure}[!thb]
\begin{center}
\includegraphics[scale=0.575,clip=true]{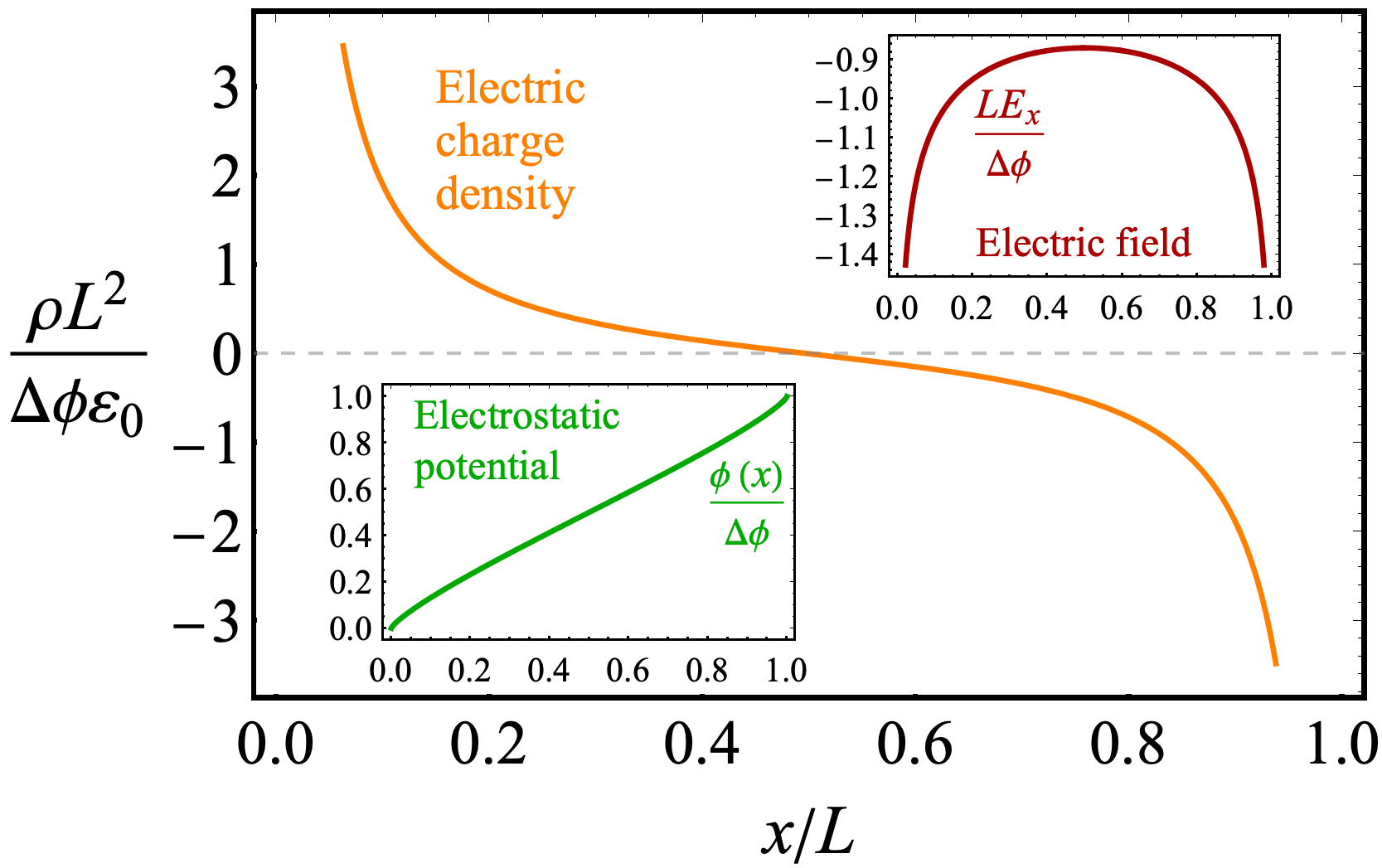}
\end{center}
\caption{The anomalous conformal screening inside a semi\-metal slab of the length $L$, characterized by a large ($\nu=1/5$) conformal screening exponent~\eq{eq:nu}. The electrostatic potential $\phi$, the electromagnetic field $E$, and the electric charge density $\rho$ are given in Eq.~\eq{eq:sol:plus} and shown above in units of potential difference $\Delta \phi$ applied to the opposite boundaries~\eq{eq:phi:bc}.}
\label{fig:examples}
\end{figure}

\begin{figure}[!thb]
\begin{center}
\vskip 3mm
\includegraphics[scale=0.525,clip=true]{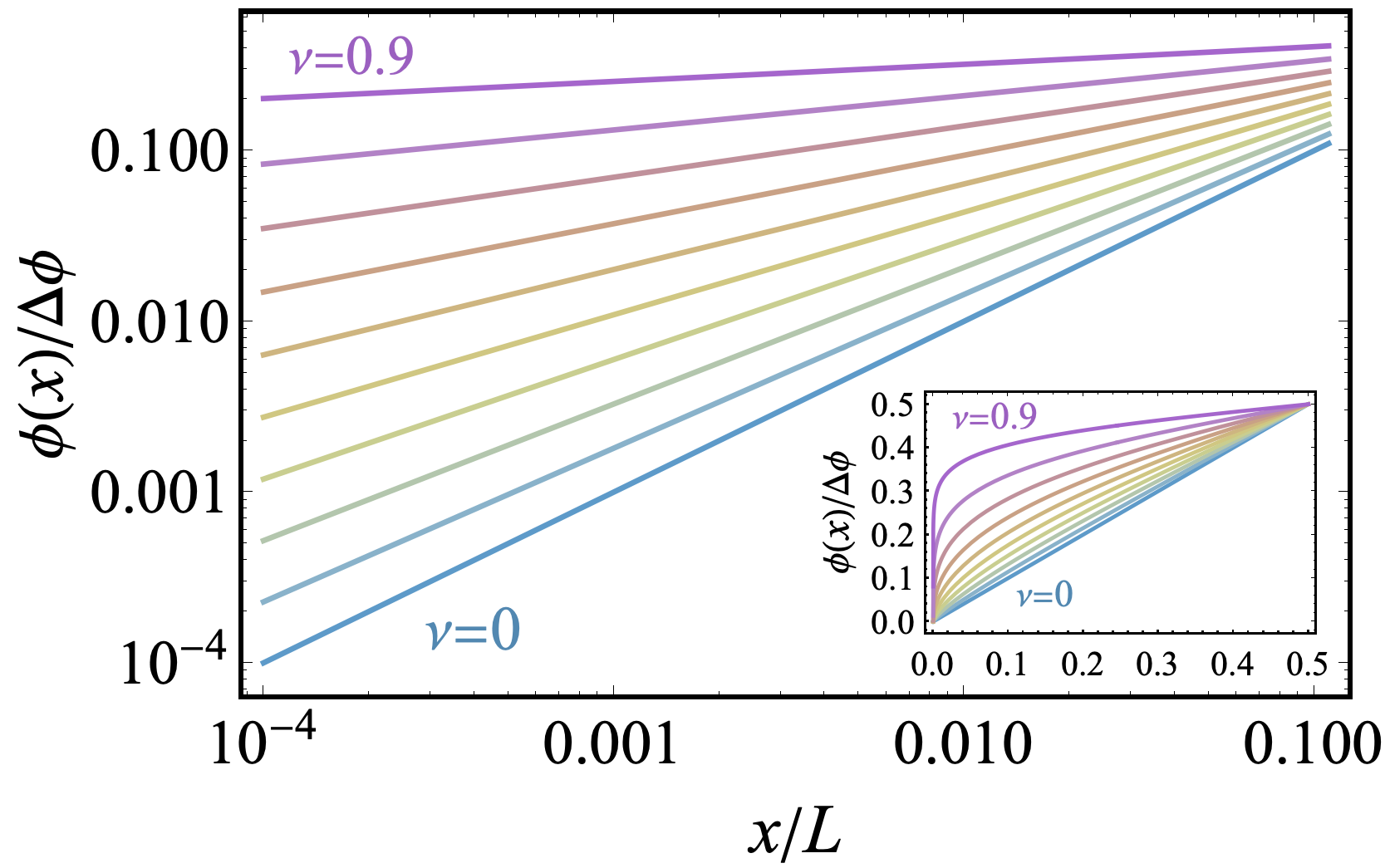}
\end{center}
\vskip -2mm 
\caption{The log-log plot of the expected electrostatic potential~\eq{eq:phi:sol:2}  due to the conformal screening near the border of the crystal for a set of the conformal screening exponents~\eq{eq:nu} $\nu=0, 0.1, \dots 0.9$ running from the bottom to the top.}
\label{fig:scaling}
\end{figure}

The static electric field inside the bulk of the semimetal is determined the conformal screening exponent~$\nu$ proportional to the beta function~$\beta_e$~\eq{eq:nu:vF}. Near the boundaries, $x \to 0,L$, the bulk fields~\eq{eq:sol:plus} reproduce the polynomial screening behaviour~\eq{eq:sol:minus}. We illustrate the conformal screening solutions~\eq{eq:sol:plus} in Fig.~\ref{fig:examples}.
\begin{figure}[!thb]
\begin{center}
\vskip 3mm
\includegraphics[scale=0.2,clip=true]{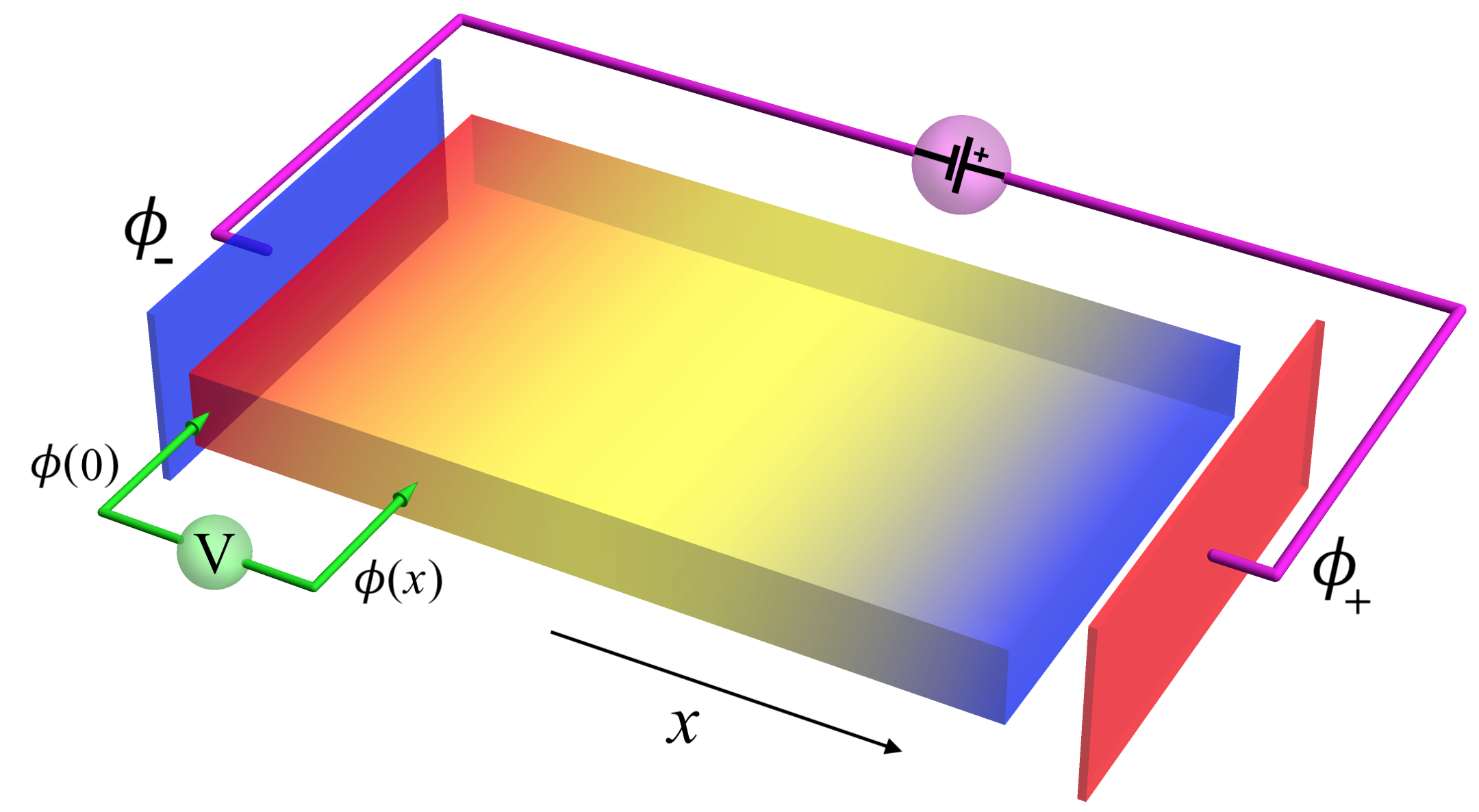}
\end{center}
\vskip -2mm 
\caption{The experimental setup to probe the anomalous conformal screening. The DC voltage source creates the difference $ \Delta \phi = \phi_+ - \phi_-$ in the electrostatic potentials at the opposite boundaries of the semimetal crystal. The behavior of the local potential~\eq{eq:sol:plus} inside the bulk, $V \equiv \phi(x)$, is determined by the conformal screening exponent~\eq{eq:nu:vF}.}
\label{fig:scheme}
\end{figure}
\paragraph{Experimental accessibility of the conformal screening.-} The anomalous conformal screening effect may be subjected to a direct experimental test. The proposed setup is depicted in Fig.~\ref{fig:scheme}. We consider a Dirac semimetal at zero temperature and zero chemical potential with a clean surface. It is sufficient to apply a potential difference to the opposite boundaries of the semimetal slab and to measure the spatial profile of the electrostatic potential difference $\phi(x)$ with respect to one of the boundaries as shown in  Fig.~\ref{fig:scheme}.  The predicted  near-boundary polynomial screening behavior~\eq{eq:sol:minus} associated with the conformal anomaly is shown in Fig.~\ref{fig:scaling}. 
In real materials the effect can be affected by standard screening due to  a finite density of states at the Fermi energy coming from impurities, departure of the the Fermi surface from the Dirac cone tip, or finite temperature effects. Moreover the continuum conformal model ceases to be valid at energies where the band bending is not negligible 
or when other bands cross the Fermi energy. Finite temperature or density induce an exponential screening characterized by a Debye or Thomas-Fermi length $\lambda_D$. The algebraic decay proposed in this work will be observable at distances from the edge smaller than $\lambda_D$ that in a Dirac semimetal is given by~\cite{WS16}:
\beqn
\lambda_D=\sqrt{\frac{\pi}{2\alpha_{WSM}k_F^2}}.
\eeqn

Moreover there is a minimal distance below which the high energy effects of the band bending will make the model invalid. The optimal experimental conditions  will then be low temperatures and materials with robust linear bands and Fermi surface close to the Dirac crossing. 
Among the symmetry-protected Dirac semimetals available at present, the best candidate should be $Cd_3As_2$ \cite{Letal14b,NX14}. Its band remains linear up to 500 meV what gives a lower bound of 0.4 $\mu$m.  The  Fermi energy  is often pined at the Dirac cone tip and can be easily tuned to match it otherwise \cite{Letal14b}. Its high Fermi velocity $v_F\sim 1,5\times 10^6 m/s$ and high dielectric constant ($\epsilon\sim 36$ \cite{JAC71} ) gives a value of $\alpha_{WSM}\sim 0.05$ and the density below 50 K is $n\sim 2\cdot 10^{16} cm^{-3}$ \cite{LH15} what gives a large  screening length length at low temperatures. 
Band structure calculations show other potential candidates with even better characteristics as  $RhSb_3$\footnote{We thank Maia G. Vergniory for suggesting these materials to~us.}. Details on the band structure of topological materials can be found in \cite{BEetal17,VEetal19}.

The surface of a real crystal may host static immobile electric charges. These built-in electrostatic defects could induce local, spatially alternating electric field in bulk of the crystal which could interfere with the effect of the conformal screening. We expect that the influence of the surface impurities may be statistically depreciated by multiple measuring the electrostatic potential at different bulk points located at a fixed distance to the crystal edge.

We expect that the proposed experiment may shed light on the polynomial screening mechanism associated with the conformal anomaly and, indirectly, with the Schwinger pair production of the quasiparticles. In addition, the experiment may provide us with direct experimental access to the value of the beta function associated with the renormalization of the fine structure constant of the fermionic quasiparticles.

\paragraph{Acknowledgements.-} We thank Qiang Li for useful conversations, and Maia G. Vergniory for providing details on the materials. We thank Dr. Rong-Xin Miao for bringing Ref.~\cite{McAvity:1990we} to our attention. The authors gratefully acknowledge financial support through Spanish MECD grant FIS2014-57432-P, the Comunidad de Madrid MAD2D-CM Program (S2013/MIT-3007),  Grant 3.6261.2017/8.9 of the Ministry of Science and Higher Education of Russia and Spanish--French mobility project PIC2016FR6/PICS07480.

\end{document}